# An MPI-based parallel genetic algorithm for multiple geographical feature label placement based on the hybrid of fixed-sliding models

Mohammad Naser Lessani[a], Zhenlong Li[a*], Jiqiu Deng[b*], and Zhiyong Guo[b]

[a] *Geoinformation and Big Data Research Laboratory, Department of Geography, University of South Carolina, Columbia, USA;*

[b] *School of Geosciences and Info-physics, Central South University, Changsha, China*

[*]Corresponding authors: Zhenlong Li (zhenlong@sc.edu), Jiqiu Deng (csugis@csu.edu.cn)

# An MPI-based parallel genetic algorithm for multiple geographical feature label placement based on the hybrid of fixed-sliding models

**Abstract**

Multiple geographical feature label placement (MGFLP) has been a fundamental problem in geographic information visualization for decades. The nature of label positioning is proven an NP-hard problem, where the complexity of such a problem is directly influenced by the volume of input datasets. Advances in computer technology and robust approaches have addressed the problem of labeling. However, what is less considered in recent studies is the computational complexity of MGFLP, which significantly decreases the adoptability of those recently introduced approaches. In this study, an MPI parallel genetic algorithm is proposed for MGFLP based on a hybrid of fixed position model and sliding model to label fixed-types of geographical features. To evaluate the quality of label placement, a quality function is defined based on four quality metrics, label-feature conflict, label-label conflict, label ambiguity factor, and label position priority for points and polygons. Experimental results reveal that the proposed algorithm significantly reduced the overall score of the quality function and the computational time of label placement compared to the previous studies. The algorithm achieves a result in less than one minute with 6 label-feature conflicts, while Parallel-MS (Lessani et al., 2021) obtains the result in more than 20 minutes with 12 label-feature conflicts for the same dataset.

## 1    Introduction

Automatic multiple geographical feature label placement is a complex computer graphic visualization problem (Harrie et al., 2022). Common standardizations are that labels be free of conflict and clearly represent their associated features. Geographical features commonly fall into three classes, including point, line, and area features. Generally, the procedure of map labeling consists of three main steps, generate a set of candidate positions for each feature, evaluate them based on their orientation and label feature association, and then select the fittest position as identified by the evaluation function (Chirié, 2000).

The nature of label positioning is proven a non-deterministic polynomial-time hard (NP-hard), where the time complexity of such a problem is directly influenced by the volume of input datasets, and its computational time grows exponentially (Christensen et al., 1995; Formann & Wagner, 1991). Major attention has been paid to addressing the problem of label placement of each type of geographical feature independently and thoroughly reviewed. However, what researchers have known in the current research is that the computational complexity of feature label positioning was ignored in most studies, and this limitation decreases the adoptability of those approaches in most realistic scenarios. For instance, the computational complexity of point label placement grows exponentially if the input features increase, although most studies were applied in small samplings. Similarly, the time complexity of line features is investigated in detail (Gemsa et al., 2020) and confirmed this problem is NP-hard in general. Without considering extensive quality factors, this problem of line label placement is solvable with $(n^3)$ execution time (Gemsa et al., 2020). Likewise, area features have been commonly regarded as points or line features based on their complex structure, which indicates label placement of polygons is challenging as point and line features.

The computation complexity of map labeling further compounds when maps contain multiple geographical features (point, line, and area) to be labeled simultaneously. However, less focus has been given to addressing the problem of multi-typed features as a cross-layer. Since the problem of label positioning is inherently NP-hard, and, to date, no algorithm has claimed to achieve an optimal solution for this problem in general. Therefore, optimization strategies are a welcomed method if they are designed comprehensively. More recently, hybrid optimization of discrete differentials and the genetic algorithm was introduced (Lu et al., 2019). Then, the idea of two degrees of freedom spaces was put forward to extend the notion of multiple candidate positions beyond point features. The introduced algorithm generates a set of candidate positions for mixed-types of features as cross-layer based on two degrees of freedom spaces (Deng et al., 2021). The above-described algorithms enhanced the quality of map visualization to a certain extent based on extensive quality functions. However, their common challenge is the computational time ranges from hours to days only for a small number of features. Consequently, these algorithms are ineffective for large-scale datasets. On the basis of DDEGA and DDEGA-NM algorithms, (Lessani et al., 2021) put forward a parallel algorithm (Parallel-MS) to address the computational time of those algorithms. Compared to the previous studies, this algorithm not only reduced the running time significantly but also achieved high-quality map labeling. Yet, its bottleneck is unaddressed, including it is only applied in spares datasets, and its computation time is unsatisfactory as the execution time for 71 multiple geographical features is over 20 and 225 minutes in parallel and serial versions, respectively, with 10000 iterations. Hence, the running time of these proposed algorithms is still a critical point besides label placement quality.

Therefore, despite fundamental improvements in multiple geographical feature label placements by the abovementioned algorithms, major challenges still exist, including computation time for large datasets and less effective labeling of multiple geographical features on maps with dense features. Improvement of the optimization algorithm is required further to initialize the best solution in the first iteration cycle and to pick the fittest chromosomes during the iteration cycle (Deng et al., 2021; Lessani et al., 2021). In the face of these challenges, there is an urgent need for new methods to be employed aiming to address the existing challenges.

In order to fulfill the knowledge gaps, this paper aims to design a comprehensive algorithm regarding the enhancement of candidate position generation and develop a fast optimization algorithm to find satisfactory results from a large number of potential solutions for multiple geographical features. To further speedup the algorithm, the notion of parallel processing is considered. In parallel computation, it is extremely important how to assign equal tasks to each CPU to optimize its performance. Overall, the contribution of this research is detailed below:

- A hybrid of the fixed position model and sliding model is regarded to reduce the number of label-label and label-feature conflicts. The sliding scheme is mainly utilized for point label placement, but here it is used at the final stage of automatic label placement, intending to move the labels that conflict with features regardless of feature type.
- The ambiguity metric is taken into consideration for all types of features. The orientation of a label with its corresponding feature is a key factor that shows the association of labels with their corresponding features.
- Position priority is applied for multiple geographical features, which enhances the readability of maps.
- A parallel genetic algorithm is introduced based on Message Passing Interface (MPI). In addition, the first population is initialized according to the lowest score of positions, which begins the iteration with a better solution.

The finding results reveal that our algorithm not only improved the computation time but also enhanced the quality of label placement. For instance, the developed algorithm labels 71 mixed-types of features in less than one minute in the serial version, and the overall number of label-feature conflicts is 6. By contrast, the running time of (Lessani et al., 2021) algorithm is 225 minutes in the serial version and 20 minutes in parallel for the same dataset with 12 label-feature conflicts. Moreover, the largest dataset includes 144 multiple geographical features in these studies (Deng et al., 2021; Lessani et al., 2021). Instead, the largest dataset in this study contains 3118 features, which are all counties in the US Contiguous States, railroads, and metropolitan cities in the US.

## 2      Related Work

*2.1 Geographical Feature Label Placement*

Multiple geographical features are the combination of point, line, and area features, and understanding the fundamental principles of each type of feature are substantially important to designing extensive algorithms for mixed types of features. Cartographers have thoroughly studied different feature types and made significant achievements with the ultimate goal of improving information visualization on maps.

Point feature label placement (PFLP) refers to the label positioning of cities, sewers, wells, etc., on the map. PFLP has gained more attention compared to other classes of features, and thorough improvements have been accomplished. In contrast, unprecedented challenges have increased alongside developments. Two models of point label placement are common, which are fixed-position and sliding models. The objectives of both approaches are identical, minimizing the overall label-feature conflict and enhancing information visualization (Guerine et al., 2019). A set of candidate positions are generated around point features as a potential location in a fixed-position model, where each has a certain preference on the basis of graphical visualization (see Figure 1(a)). This scheme of point labeling has been broadly welcomed by the cartographic community due to its applicability. This configuration established a great venue for research, and numerous optimization strategies were taken into account for addressing the problem of PFLP on the basis of a fixed position scheme, including genetic algorithms (Verner et al., 1997), greedy algorithms (Cravo et al., 2008), simulated annealing approaches (Zoraster, 2013), and integer programming (0-1) (Marín & Pelegrín, 2018). In addition, the concept of conflict graph, which has been widely implemented in various domains, is applied to address PFLP (Gomes et al., 2013), as shown in Figures 1(d) and (e). The objective of point labeling within this concept is to find the maximum independent number of conflict graphs. Recently a rapid method for PFLP was proposed running on GPU (Pavlovec & Cmolik, 2022). However, finding a promising solution is unlikely in this optimization primarily, but its computational time is much faster than the preceding approaches. Furthermore, only a number of these studies considered the common information visualization criteria, like label ambiguity, label priority, label-feature conflicts, and label-label conflicts. Recently, the sliding model has gained attention as well, where the labels move around the features until finding an appropriate position, as shown in Figure 1(b) (Li et al., 2016).

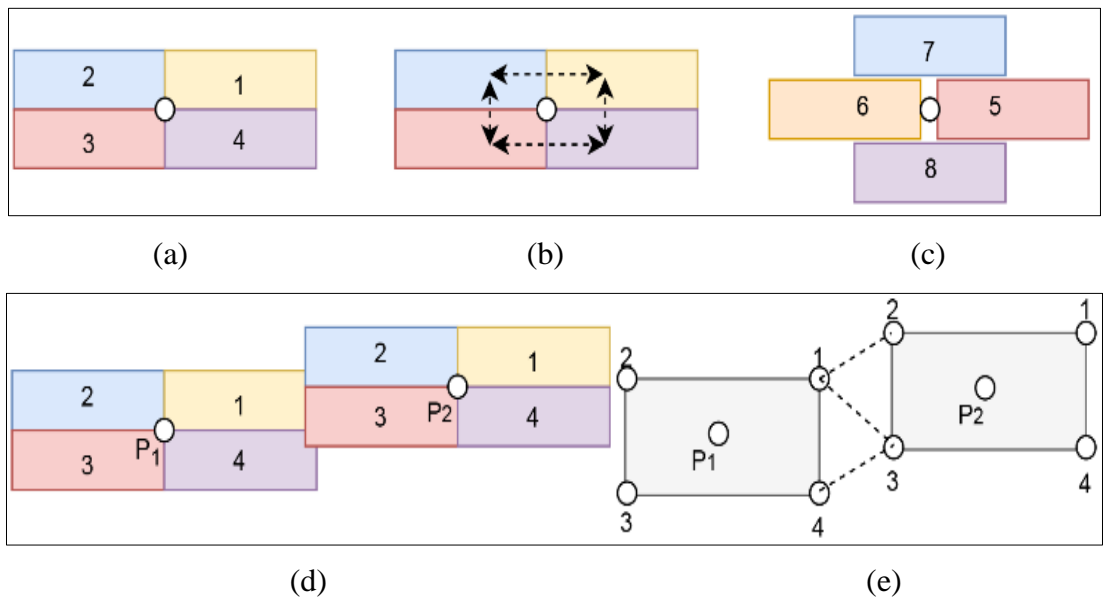

Figure 1. Illustrate the instances of point labeling models and conflict graph: (a) shows a point feature with four pre-defined candidate positions; (b) shows a sliding model for point labeling; (c) shows the scheme with eight pre-defined candidate positions plus with condition (a); (d) shows the condition of label-label conflict between two features; (e) present label-label conflicts according to the notion of conflict graph based on condition (d).

A broad range of studies was conducted to address the PFLP, yet less thought has been given to labeling lines and area features. Generally speaking, regarding geometrically, line features are a collection of segments representing objects on the ground that fall into a line shape. To visualize such objects clearly with their names on the map, several rules and quality criteria have been introduced according to cartographic standardization, (Chirié, 2000) is one of the pioneer researchers gathered a set of rules by interviewing cartographers. The criteria in that research are mainly classified into three classes: labels are positioned within and parallel to the line shape; roads intersection between junctions must be vividly identified; the label conflicts of two roads are prohibited. What is more, he put forward a mathematical definition to label a single road, and a heuristic algorithm was introduced for sequentially labeling all roads on the map. Each of these rules is extended and well-researched. Additionally, one more factor was considered for line features in that research; for example, the labels alongside the lines is in higher preference (Niedermann & Haunert, 2018). The authors further gave a layout of the metro map that contained several metro lines. First, a set of general models were defined to label metro maps and then generated a set of discrete candidate positions similar to point features. Moreover, the idea of graphs was also applied by presenting the label positions as the vertices and the road network as edges. A recent effort was made based on an algorithmic approach, which was considered from the perspective of mathematical description to maximize the number of sections being labeled (Gemsa et al., 2020). The computational complexity of their algorithm is bound to $O(n^3)$ [4]. Furthermore, the study considered the concept of grid-shaped for road networks, (Neyer & Frank, 2000) made a specific algorithm with high efficiency to investigate such kind of grid-shaped approaches if possible. Even though researchers have witnessed many attempts to find best solutions to this problem recently, it is proven that determining a label for each road in the road networks without label-label conflict is NP-complete, which is very unlikely to come up with a polynomial worst-case running time. Figures 2(a) and (b) are the illustrations of line labeling.

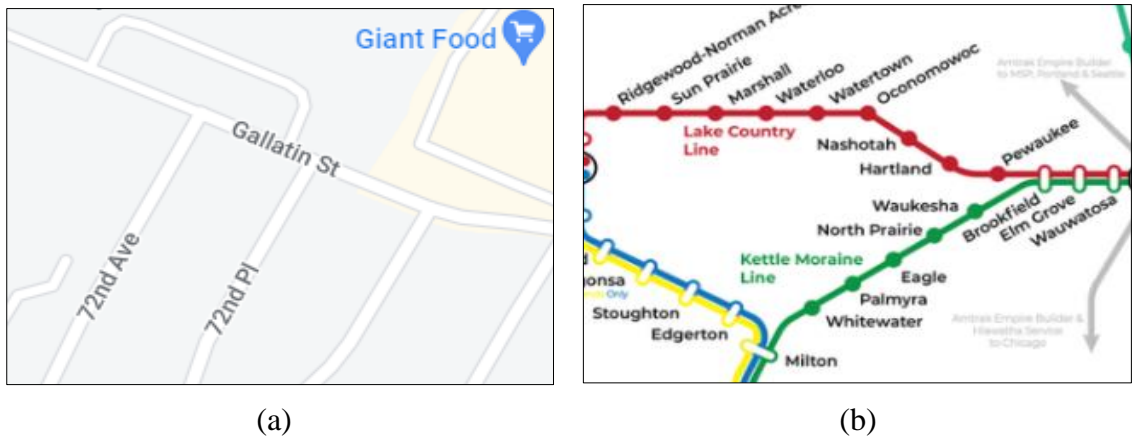

(a)                                     (b)

Figure 2. Representation of road networks labeling: (a) shows the labels within roads; (b) shows a railway network with labels around the line features.

Besides point and line features, another important type of feature is the area (polygon). Area features usually are bordered with liner objects and other types of objects, including other area features or points. This degree of complexity makes assigning an optimal place for their labels more challenging. Labels of area features are positioned within or outside features. In the case of placing the label inside the features, two major notions are regarded. Firstly, applying the idea of line labeling to generate a set of label positions within the polygon based on skeleton lines if the polygon is large enough. Second, generating a number of discrete positions within the polygon (Barrault, 2001; Li et al., 2020; Pokonieczny & Borkowska, 2019; Rylov & Reimer, 2017). Otherwise, the candidate positions are generated outside the area feature if the polygon lacks enough inner space (Rylov & Reimer, 2017). In terms of computational complexity, the complexity of area features also increases exponentially by adding more features when it is determined as a point or line. Figure 3. shows the instances of label positioning of area features within and outside features.

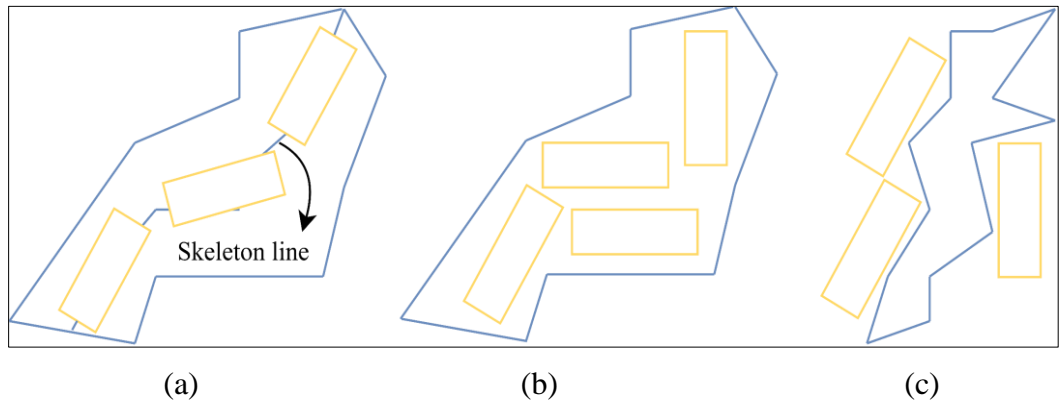

Figure 3. Representation of label placement of area features: (a) presents candidate generation based on skeleton line within the area; (b) presents candidate generation based on the discrete model; (c) presents external label placement scenario.

*2.2 Parallel Programming*

Several parallel programming models were put forward in the early age of computer science. In comparison, Message Passing Interface (MPI) was most desirable after its release in the 1990s (Clarke et al., 1994; Gropp et al., 1996). MPI standard was initially designed for the distributed memory system, and the advantage of multi-threading that is originally based on shared memory was neglected (Zhou & Tóth, 2020). However, due to the adaptability of MPI and its friendly environment, the software vendor started to add more features that enabled the model to be applied in shared memory systems, too. Also, as the domination of parallel computation became evident shortly after its release, researchers considered accelerating the usage of various models of parallel programming by combing them, for instance, hybrid parallel programming, which is the combination of MPI and OpenMP models (Rabenseifner et al., 2009).

In MPI parallel programming model, to run the algorithm concurrently on multiple processors/machines, the processors necessarily demand communication with each other through MPI library routines. Basically, the program is launched independently by a set of processors, and each one has an identical process. These generated processors execute the same or different program code and instructions.

Generally, a parallel program gains its highest performance when every MPI is mapped on a disjoined processor, and the amount of communication is minimized during execution. In that sense, to make communication easier among initialized processes, the MPI library supplies functionality for communication and synchronization. One of the basic functions is point-to-point communication, which is particularly helpful in implementing an irregular communication pattern. Furthermore, point-to-point communication includes several modes of communication, including synchronized, asynchronous, and buffered. In the form of asynchronous communication, programmers are able to perform different types of MPI operations (Espínola et al., 2017). To facilitate exchanging message among processors, MPI interference provides MPI_Send and MPI_Recv, where MPI_Send sends information to the specified processors according to the instruction and the destination processor receives the message by MPI_Recv call (Bernholdt et al., 2018). Detailed studies are carried out on these researches regarding various types of communication modes (Dosanjh et al., 2021; Hoefler et al., 2007).

## 3 Methodology

### 3.1 Candidate position generation

The scheme of candidate position generation varies according to the type of features on the map, as mentioned in the literature section. In this study, the fixed position model is considered to generate a set of discrete candidate positions for point features in the first phase of the algorithm. Also, the concept for producing candidate locations for the area and line features is similar, which is based on a skeleton line. This set of candidate positions is limited however for each feature in practice.

To generate candidate locations around point features, we considered producing 24 candidates on three layers neighboring to the point, as illustrated in Figure 4; and positions are scored according to the cartographic standardization. As the number of candidate locations increases, the possibility of labels being placed free of conflict is improving. As presented in Figure 4. the total number of candidate positions is 24, and this enhances the probability of labels being positioned more clearly (Deng et al., 2021). The following equation is established to generate candidate positions around the point.

$$x_i = x_p + R * \cos(\beta) \quad , y_i = y_p + R * \sin(\beta) \tag{1}$$

Where $(x_p, y_p)$ is the coordinate of point $(i)$, and $(x_i, y_i)$ is the coordinate of the closest point of the label box to the point $(i)$, $(R)$ is the distance between the point feature and the closest point of the label box. The value of $(R)$ varies based on the layer distance from the point features, as can be observed in Figure 4. $(\beta)$ is the angle between the closest point of the label box and the point and axis $(X)$.

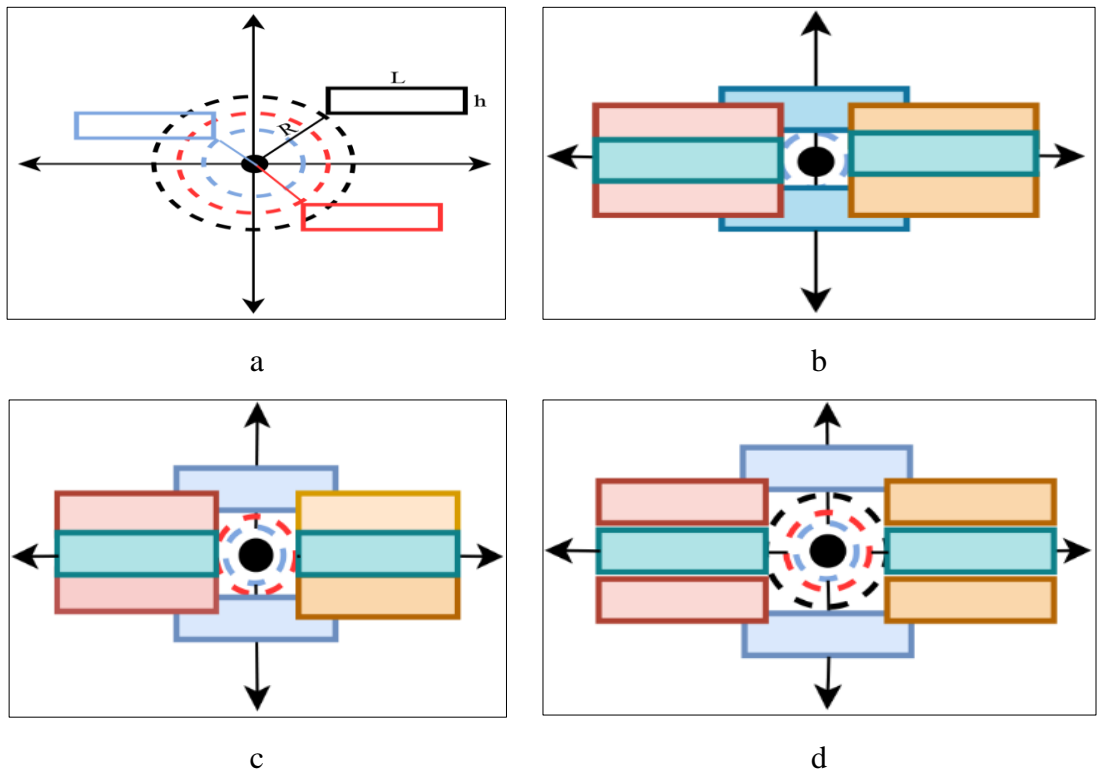

Figure 4. Illustration of candidate position generation for point feature: (a) shows three layers around the point feature, which defines the space for candidate locations; (b) 8 locations on the first, similarly the locations are generated on the second and third layer (c and d).

Candidate positions for area features are produced based on skeleton lines in this research. Thus, the accuracy of candidate locations for polygons depends on how the skeleton lines are generated because initiating a skeleton line can be challenging in some cases, particularly when the shape of a polygon is complex. In order to initialize the skeleton line for polygons, the introduced algorithm divides the input polygon into 8 equal segments according to polygon shape by a parallel line, and then the midpoint of these lines is connected together. If a parallel line has multiple intersection points with the polygon, the line with the longest distance is considered the skeleton point. Also, these intersection points are regarded as the centre of the label box. Once the approximate skeleton line is generated, two parallel skeleton lines are produced, one above and another below the initial skeleton line. The preference of location is horizontal, along the skeleton line, and vertical for the polygon, respectively. A real example of how labels are generated on a skeleton line is illustrated in Figure 5(d-f); as it can be seen, the angle for all three labels on the right side is changed from horizontal due to label-feature conflict. Yet, a small portion of labels on the left side are in conflict with the boundary of the polygon, and this can be removed by pushing to the right side in the sliding stage if one of these locations is selected in the optimization phase. By the same token, the labels are generated for polylines. However, the location parallel to the line has higher preference than horizontal and vertical polyline.

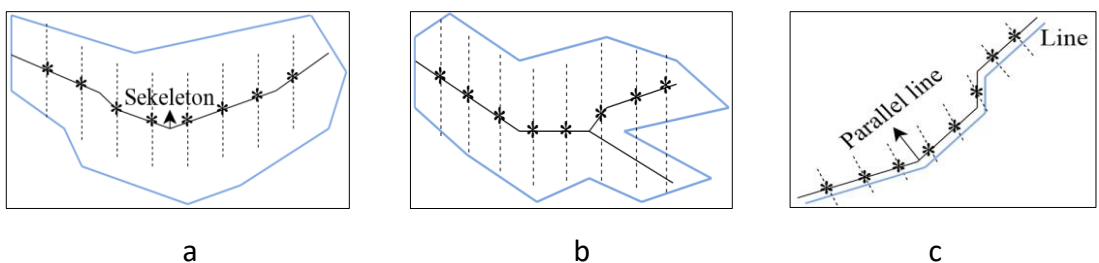

a                      b                      c

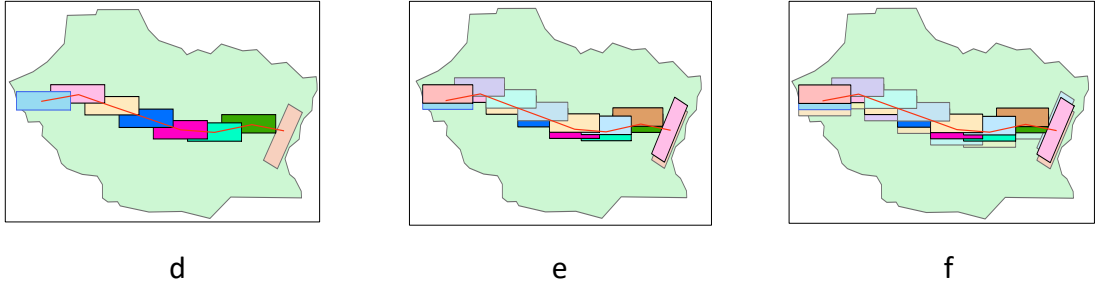

d             e             f

Figure 5. (a, b) Demonstration of how skeleton lines are generated for polygons, and each stark shows the centroid of label box; (c) presents candidate locations for polyline, and the label box centre is marked with stark; (d) presents a real polygon with 8 candidate locations on the first layer, and the red line shows the generated skeleton line; (e, f) show 8 candidate positions on second and third layers on top of the first layer.

Finally, after label box centre construction, identifying the intersection of generated label box with features are necessary to evaluate label-feature conflict. On this basis, the minimum bounding rectangle box (MBRB) is created based on the coordinate of the label box centre, and the equation is defined in Equations (2) and (3) to construct the coordinate of MBRB. The $(x)$ coordinate is constructed by $(p_{x\_i})$, and $(p_{y\_i})$ extracts $(y)$ coordinate, $(x_c, y_c)$ is the coordinate of the label box centre, $(\theta)$ stands for label box angle (horizontal, parallel to the skeleton or line feature, or vertical), and the length and height of the label box are shown by $(L)$ and $(h)$, respectively.

$$p_{x\_i} = (x_c \pm \frac{L}{2} * \cos(\theta) \pm \frac{h}{2} * \sin(\theta)) \quad (2)$$

$$p_{y\_i} = (y_c \pm \frac{L}{2} * \sin(\theta) \pm \frac{h}{2} * \cos(\theta)) \quad (3)$$

Generating 24 candidate locations is nonessential for those polygons that are free of intersection. This improves the computational complexity of map label placement since the problem of label positioning is an NP-hard problem where the computation time of an algorithm grows exponentially. For example, if the input feature $(N = 10)$ and the number of candidate positions is 24 for every feature, the set of solutions is $(24^{10} \approx 63 *$

$10^{12}$), it is extremely challenging to explore the optimal result from such a huge number. On the contrary, if it is reduced from 10 to 5, the overall solutions would be ($24^5 \approx 8 * 10^6$). On the basis of this argument, one position is generated for polygons that are without intersection based on the proposed algorithm in order to ease the complexity of the problem and save computation time.

## 3.2 Quality function

In text labeling, label-label conflict, label-feature conflict, label ambiguity, and label position priority are critically important intending to enhance the quality of information visualization on the map. A comprehensive quality function is formulated in this study, which contains general graphical information visualization criteria for multiple geographical features.

$$\rho_2(\gamma) = \sum_{i}^{4} S_i \tag{4}$$

In Equation (4), 4 is the number of quality factors ($S_i$) represents each quality metric. The objective is to minimize the $\rho_2(\gamma)$ function to ensure that the best positions are quantitatively evaluated. The instance where two or more label boxes overlap each other is termed label-label overlap, which leads to leaving some features without labels, as defined in Equation (5). In contrast, when a label overlaps with one or a set of features on the map, label-feature conflict is described, and this causes a reduction in the readability of labels.

$$S_1 = \sum_{i=1}^{Q-1} \sum_{j=i+1}^{Q} x_{i,j} * 9 \tag{5}$$

Where ($Q$) stands for the number of features with 24 candidate positions, whereas in the DDEGA-NM algorithm, this value is equal to the total number of input features. If ($x_{i,j}$) is greater than zero, then label ($i$) overlaps with label ($j$); otherwise, label ($i$) is free of overlap, and the value of 9 is a constant value because the score of label-label conflict is set higher than other quality metrics.

$$\mu_1 = \begin{cases} 0, & Without\ feature\ conflict \\ 1, & label - line/area\ conflict \\ 99, & label - point\ conflicts \end{cases} \quad (6)$$

$$S_2 = \sum_{i=1}^{N} \mu_{i(1)} \quad (7)$$

Equation (7) is defined for label-feature conflict. Here, ($N$) stands for the input features, and three values are assigned for ($\mu_1$) based on the instance of label-feature conflicts.

$$\mu_2 = \frac{D - D_{min}}{D_{max} - D_{min}} \quad (8)$$

$$S_3 = \sum_{i=1}^{N} \mu_{i(2)} \quad (9)$$

Equation (9) is established to evaluate the ambiguity of generated labels with the corresponding feature. In the DDEGA-NM algorithm, this metric was only considered for polyline; however, we regarded all three types of features. Where ($D$) is the distance from the centre of the label box to the point-for-point feature, the midpoint of a polyline, and the skeleton line for a polygon, ($D_{min}$) is the distance from the feature to the first layer, and ($D_{max}$) presents the distance from the corresponding feature to the third layer.

$$\mu_{Q1} = \begin{cases} 0.25, & \beta \in [0°, 90°) \\ 0.5, & \beta \in [90°, 180°) \\ 0.75, & \beta \in [180°, 270°) \\ 1.0, & \beta \in [270°, 360°) \end{cases} \quad (10)$$

$$\mu_{Q3} = \begin{cases} 0.75, & Vertical\ label \\ 0.5, & Along\ the\ skeleton\ line \\ 0.25, & Horizontal\ label \end{cases} \quad (11)$$

$$S_4 = \sum_{i=1}^{Q_1} \mu_{i(Q1)} + \sum_{i=1}^{Q_3} \mu_{i(Q3)} \quad (12)$$

Equation (12) is defined to assess label position priority for points and polygons,

but this quality factor was only regarded for point features in the previous study. The steps for generating candidate positions are shown in Algorithm 1 on the supplementary material.

*3.3 Optimization Algorithm*

After generating the candidate positions, the optimization phase of the algorithm begins. The optimization phase of the developed algorithm is on the basis of discrete differential and genetic algorithm (Lu et al., 2019), where both of them is combined to enhance the performance of the optimization step (Katoch et al., 2020). The evolutionary algorithm starts with an initial random solution, termed as the initial population. The selection of the first population is highly critical and influences the remaining iterations. In this paper, the total set of combinatorial solutions is ($C_{pos} = 24^{F_{24}}$), where ($F_{24}$) is the number of features with (24) candidate locations. The locations are sorted in ascending order based on the lowest score for each feature first, 100 populations are initiated in the first iteration then ($p_1, p_2, p_3, \ldots p_{np=100}$), among them one chromosome includes positions with the lowest score value regardless of label-label conflict; this avoids random population initialization. The fitness function of the optimization algorithm is expressed in Equation (13).

$$f_{(x)}^n = \sum_{i=1}^{4} \sum_{j=1}^{np} S_{i(p_j)} * W_{i(p_j)} \tag{13}$$

Equation (13) is established as the fitness function, and the objective is to minimize the function $f(x)$. The weight of label-label and label-feature conflicts are set greater compared to other metrics since these factors are critically important for map label placement. In this equation, ($S_i$) presents the score of each quality factor and ($W_i$) stands for the weight value of each corresponding quality metric. ($n$), ($np$) and ($p_j$) stand for the number of iterations, the population size, and the population ($j$), respectively.

70% of the population is initialized by genetic algorithm and 30% based on Discrete Differential Evolution (DDE) algorithm to benefit from both evolutionary algorithms. 70% percent of the population is generated according to a genetic algorithm using Equation (14), in which ($x$) is a random number within the range of the population, ($k$) and ($j$) stand for populations ($k\ and\ j$) within the entire population ($np$). It demonstrates that child ($P_{GA}$) is produced from parents ($k\ and\ j$) based on the crossover point ($x$). Three vectors are randomly selected from the entire previous population while generating 30% of the population using the DDE algorithm, and the difference between the two of them is scaled by a scale factor ($F$), as shown in Equation (15). Equation (16) is defined to depict the mutation operator; again, ($x$) is a random number with a range of candidate positions, and ($y$) presents a feature within the input features ($N$). $p_{j(y)}$ means the candidate location of the feature ($y$) is modified in population ($p_j$) according to the defined instances.

$$P_{GA} = p_j(x) + p_k(x);\ x \in [0, np];\ k \neq j;\ k, j \in [p_1, p_2, p_3, \dots p_{np=100}] \quad (14)$$

$$P_{DDE} = \left(X_{r1}(t) + F * (X_{r2}(t) - X_{r3}(t))\right) \quad (15)$$

$$M = \begin{cases} x \in [0,22], & y \in [0, N] \\ p_{j(y)}(x+1), & if\ p_{j(y)} > x \\ p_{j(y)}(x-1), & if\ p_{j(y)} < x \end{cases} \quad (16)$$

The unvaried procedure continues until all the number of iterations is completed or the score value of the achieved result is lower than the specified threshold value. The selection of appropriate values for mutation and crossover rate is crucially important. We set these ratios at 0.01 and 0.5, respectively, based on experiments. The procedure of candidate positions and the optimization phase of the algorithm and genetic algorithm operator are presented in Algorithm 2 and Figure 1 on the supplementary material.

*3.4 MPI parallel processing for map label placement*

Numerous map label placement algorithms have been developed over decades aiming to enhance label placement quality. Nonetheless, less attention has been given to mixed-types of feature label positioning until recently. (Deng et al., 2021; Lu et al., 2019) studies addressed the problem of geographical features as a cross-layer, yet the computational time of these publications is critically long. Next, the Parallel-MS algorithm was introduced, mainly focusing on reducing the computation time of label placement (Lessani et al., 2021), but despite the great improvement, its result is unsatisfiable. For example, the execution time of 71 geographical features is 20 minutes in parallel and 225 minutes in sequential, and this reduces its practical application. In this paper, a new parallel algorithm is developed for label positioning based on Message Passing Interface (MPI) that significantly enhances the proposed algorithm. Efficiently applying parallel processing further accelerates data analysis; otherwise, the advantage of utilizing multiple cores/machines is insignificant.

Figure 6. illustrates how the master node and slaves are functioning, one Central Processing Unit (CPU) is determined as the master node, and the remaining CPUs are assigned as salves (workers). Automatically all existing CPUs are activated in the machine by calling MPI, and then the input features are divided equally on each CPU because workload balancing is an essential component of parallel processing, where every processor needs to have an equal number of tasks to ensure maximizing the performance of the algorithm (Ouazene et al., 2021). Next, the input features are scattered to salves by the master node. Each processor carries out identical instructions but supplies its instruction with its unique data. After the completion of candidate position generation on each CPU, the results are sent back to the master node by slaves. Then the master node classifies the features into two groups: polygons with one position and features containing

(24) candidate locations. In the second stage, the features with (24) candidate positions are broadcasted back to salves by the master node, and the optimization step begins. During the iteration cycle, the best results are shared by each CPU after (500) iterations, and this procedure continues until the stopping criteria. The completion time ($C_t$) of the algorithm is expressed in Equation (17), the set of tasks is shown by ($S_j$), and ($p_i$) presents the processing time of the task ($i$).

$$C_t = \sum_{i \in S_j} p_i \tag{17}$$

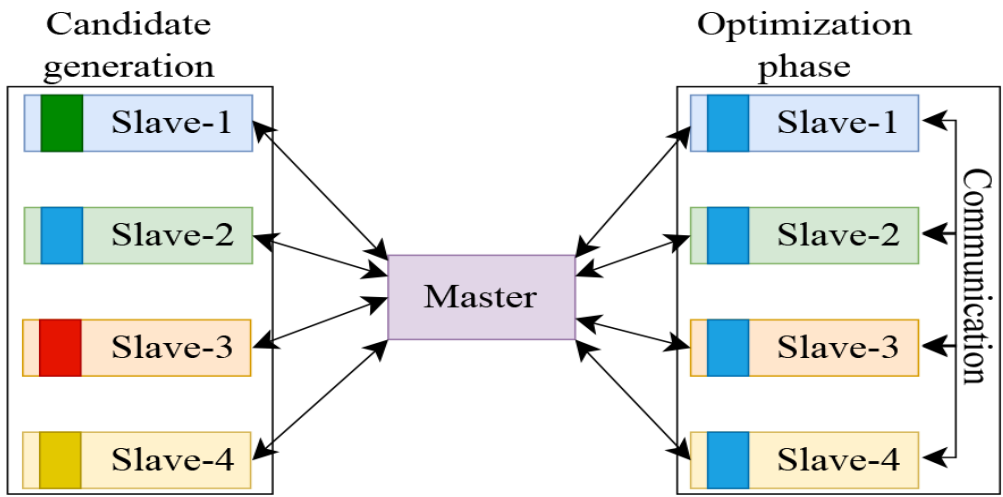

Figure 6. Representation of MPI stages in the proposed algorithm: candidate position generation, where each slave has its unique data; the input data is the same for all slaves in the optimization phase (features with 24 positions), and slaves communicate through the MPI library routine.

*3.5 Sliding the selected labels*

The sliding model is generally used for point feature label placement, where a label is moving around the feature until the label-feature conflict is removed. Yet, less attention has been paid to this scheme of label placement compared to the fixed position model. In this proposed algorithm, this model of label positioning is added on top of the fixed position scheme in order to benefit from both models, which declines the number of label-feature conflicts, especially when the input data is larger because optimization approaches are unable to explore the best position for all input features in every case. For instance, the contiguous map of the US has 3351 features in total, and among them, 1159 features have 24 positions each. Therefore, it generates $24^{1159}$ solutions that are impossible to be explored the best position with 10000 iterations from this huge number. Thus, the sliding model can play a key role in resolving the issue of label-feature conflicts after the optimization phase.

To identify label-feature conflicts after the optimization stage, the status of selected positions is investigated with the input features; if a conflict occurs, the label position moves with regard to label-label conflict. Whereas the experimental datasets include mixed-types of geographical features in this study, separate steps are considered for each type of feature in this phase. Since label conflict with points is a higher priority compared to the line and area features, the position moves around with regard to the neighboring features until the conflict is eliminated, as shown in Figure 7(a). However, if the conflict still exists, a new position is determined according to the fixed position scheme. Two instances are indeed considered for replacing the labels of points in the sliding step: 1) completely changing the location of the selected position, as shown in Figure 7(c); 2) moving the existing position to a certain direction, as shown in Figure 8(c).

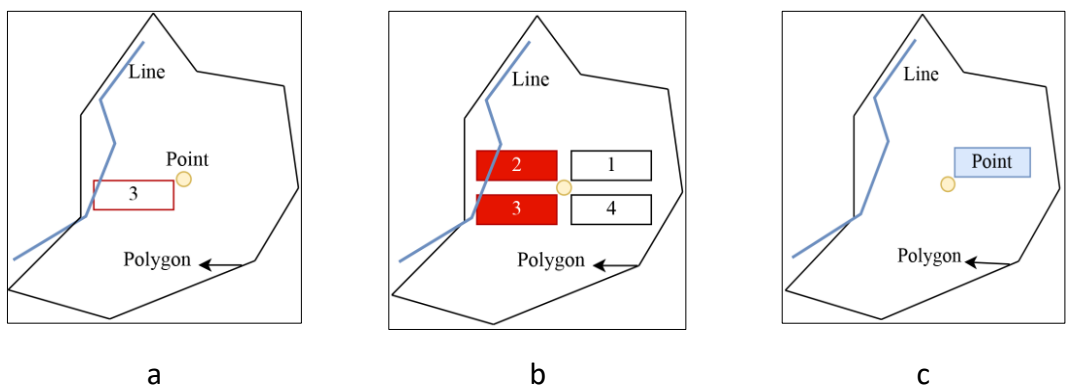

a             b             c

Figure 7. Representation of label movement for point feature for the first case scenario: (a) shows the chosen position has a conflict with a line; (b) positions with indices 1 and 4 are free of conflict; (c) the location with index one is selected for the final output.

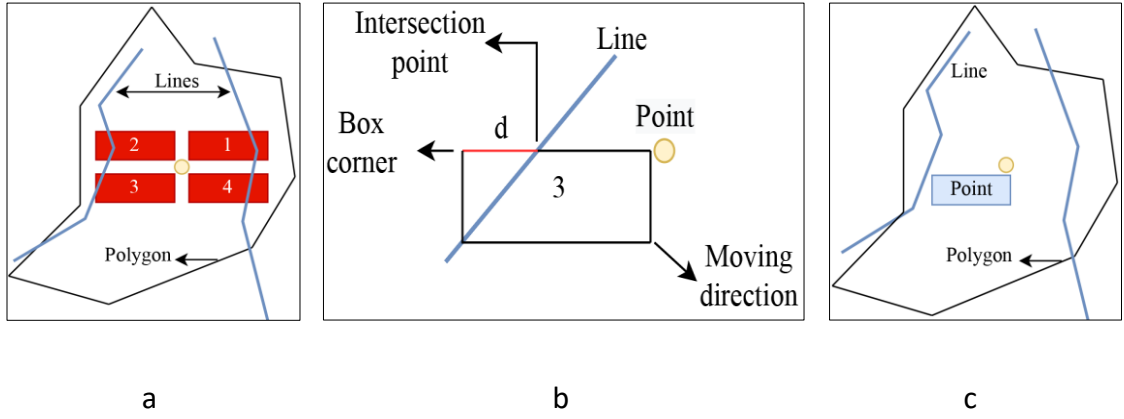

a  b  c

Figure 8. Representation of label movement for point feature for second case scenario: (a) suppose all four positions are in conflict; (b) the selected position in the optimization phase needs to move into another direction to eliminate label-feature conflict if it is possible; (c) selected position free of conflict based on moving the labels to a certain direction.

$$\sum_{i=1}^{l}\sum_{j=1}^{N} x_{i,j} = 0 \text{ or } > 0 \qquad (18)$$

$$d = \|r - b\|^2 \qquad (19)$$

$$d_{mve} = d + \varepsilon \qquad (20)$$

$$new_{pos(I)} = (p_x \pm b * \cos(\theta), p_y \pm b * \sin(\theta)) \qquad (21)$$

$$new_{pos(II)} = (x_c \pm d_{mve} * \cos(\theta), y_c \pm d_{mve} * \sin(\theta)) \qquad (22)$$

In Equation (18), ($l$) and ($N$) stand for the set of labels and input features, respectively. If the value of ($x_{i,j}$) is zero in Equation (18), the selected position is free of conflict; otherwise, conflict occurred. ($i$) stands for label position and ($j$) presents the feature. The Euclidean distance is calculated between the intersection point ($r$) and the corner of the label box ($b$) by Equation (22) for the next step. In Equation (20), $\varepsilon$ presents a constant value, and this value is added to the actual distance between ($r$ $and$ $b$) to ensure the label box is moved more than the actual distance ($d$). Equation (21) is defined for the instance of selecting a new position, as shown in Figure 8(b), where positions with indices 1 and 4 are without conflict; $p_x$ and $p_y$ is the coordinate of the point feature, and $b$ presents half of the label box height. However, if all four positions are in conflict, the algorithm moves the chosen position in a different direction in order to eliminate label-feature conflict based on Equation (22), as shown in Figure 9(c). In this Equation, $x_c$ and $y_c$ presents the coordinate of the centre of the label box, and ($\theta$) includes $[45°, 135°, 225°, 315°]$ in both Equations (18) and (19).

Candidate positions are generated along the skeleton line for area features; thus, polygons and lines are considered similar features in this step. When labels of the area and/or line conflict with other features, the chosen position moves in such a way that eliminates the conflict or identifies a better location. If label-feature conflict omission is impossible, the process of label movement for line and area features is shown in Figure 9. The process of label movement for the area and line features follows Equation (22), but the value of ($\theta$) is different (0 $or$ 90), or it's equal to the angle of line and skeleton line direction. In addition, to prevent label-label conflict while moving the labels, the overlap of the newly generated position is investigated with other neighboring labels. For instance, when the label 'Polygon' moves to a new position, its overlap is evaluated with the label 'Line' then to avoid label-label conflict.

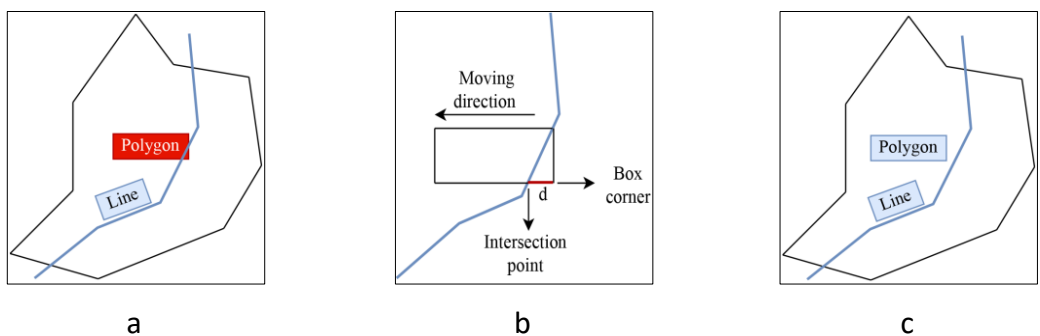

a            b            c

Figure 9. Illustration of label movement for the area and line features: (a) presents the label of the polygon in conflict with the line feature; (b) shows the direction of label movement; (c) presents the final label placemen.

## 4       Result and discussion

The developed algorithm is executed on a computer with 16 GB RAM installed, Intel Core i5-8265U CPU @1.60 GHz 1.8 GHz, and the installed windows is "windows 10 Professional × 64". The program is written in Python language under the Pycharm framework. Message Passing Interface (MPI) is used for the implementation of parallel processing with the MPICH2 version. To verify the practicality of the introduced algorithm in map label placement, four real-world datasets are selected for the experiments, and they are shown in Table 1. The first map includes 3351 multiple geographical features, and the second dataset is downloaded from OpenStreetMap (OSM) where located in Washington with the coordinates (-76.16354, 43.03704). The last two datasets are selected to evaluate the performance of the developed algorithm with previous studies since they implemented their experiments on these maps.

Table 1. Experimental datasets.

| Dataset | Area | Line | Point | Total | Location |
|---------|------|------|-------|-------|----------|
| First   | 3088 | 72   | 191   | 3351  | US counties |
| Second  | 703  | 75   | 36    | 814   | OSM (Washington) |
| Third   | 88   | 20   | 36    | 144   | Ohio, US |
| Fourth  | 39   | 17   | 15    | 71    | Washington, US |

The efficiency of the proposed algorithm is analyzed in two aspects: comparison with other algorithms in terms of computation time and visual effects. For computation efficiency, the computation time of the algorithm is compared with the other three algorithms. Regarding visual effects, the results are analyzed based on the number of label-label and label-feature conflicts and the selected position for labels. Whereas the time computation for the first dataset would be critically long using the previous studies, thus, its result is only compared with MapLex Label Engine in ArcGIS.

## 4.1 The performance of the optimization algorithm

The proposed algorithm consists of three stages, candidate position generation, optimization phase, and sliding stage. The basis of our optimization algorithm is the DDEGA algorithm (Lu et al., 2019), where discrete differential evolution and genetic algorithms are hybridized. DDEGA algorithm initializes the population randomly regardless of considering starting the iteration cycle with a better population. Conversely, the populations are initialized based on the lowest score of the candidate position in this study. This allows the algorithm to find the solution faster and avoids unnecessary iterations, as it can be observed in Figure 10 that the gap between two bars is wide. When the dataset is small, the gap between two graph lines is wide at the beginning of the iteration, whereas the difference between the two final score values is smaller. This verifies that the algorithm is able to retrieve a satisfactory result for a small dataset regardless of how the population is generated. In contrast, when the input dataset is larger, the importance of how to produce the initial population becomes more obvious, as shown in Figures 10(a) and (b). The range of potential solutions in such scenarios is huge; subsequently, the algorithm requires more iterations and computation time to get a fitting solution. If the population of the algorithm is initiated appropriately, the algorithm requires less iteration cycle and less computation time.

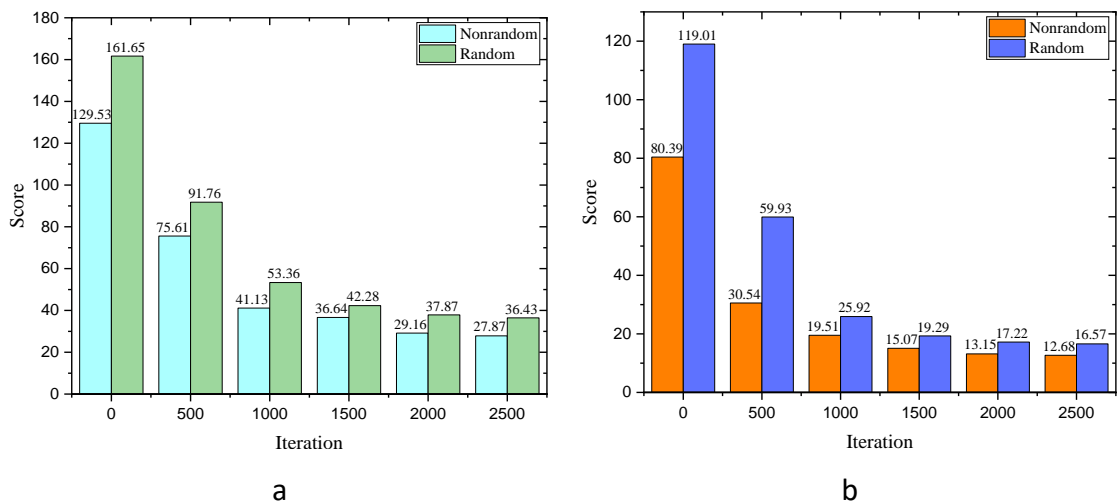

Figure 10. Visualizes the differences between random and nonrandom population initialization: (a) Ohio State dataset, (b) OpenStreetMap dataset.

Since the algorithm runs in parallel, all CPUs need to communicate during the iteration cycle in order to share their best solution with each other, as shown in Figures 11(a) and (b). This permits the algorithm to maintain diversity in the population. Otherwise, it probably leads to a problem known as premature convergence though the optimization algorithm includes a mutation operator. The probability of such an instance increases as the complexity of the problem grows. In this regard, CPUs communicate after every 500 iterations to maintain diversity in the population on top of the mutation operator, as illustrated in Figure 11. CPU1 has achieved the best solution after 500 iterations among other CPUs, as demonstrated in Figure 11(b); on this account, this chromosome is inserted into the population of other CPUs for the next iterations. Whereas after 1000 iterations, CPU2 has obtained a better solution compared to other CPUs, and its chromosome is shared with other processors thus.

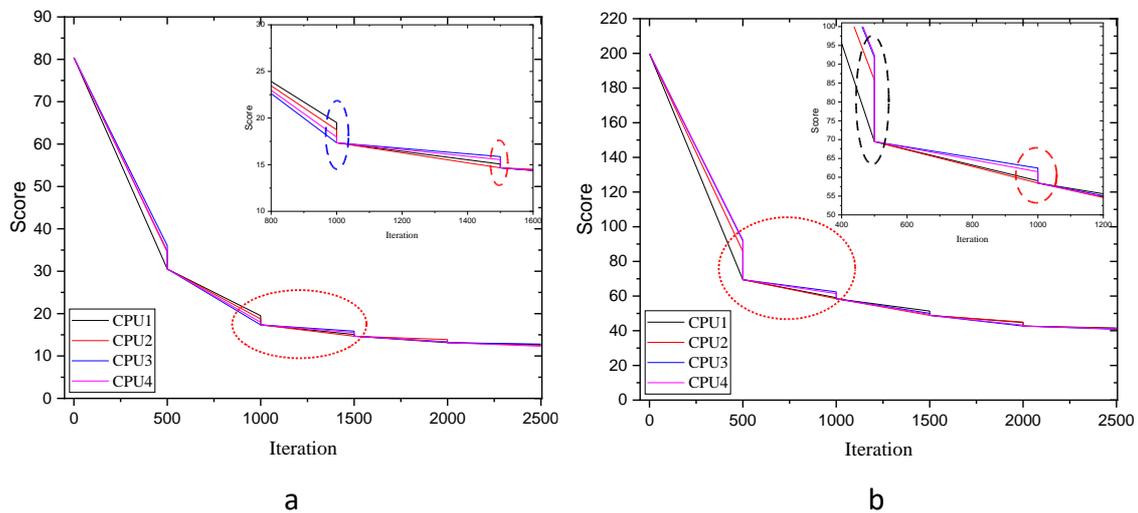

Figure 11. Visualizes sharing solutions between CPUs in the optimization stage during the iterations: (a) presents the OpenStreetMap dataset and (b) shows the first dataset.

*4.2 Sliding label position to eliminate label-feature conflict*

This section describes the result of an alternative procedure for those instances in which the optimization phase fails to remove label-feature conflict within the specified iterations. Two situations are considered in this step either label-feature conflict is eliminated, or they are placed in a better location. When the optimization phase is completed, label-feature conflict is investigated for the selected label positions. If the label location overlaps with the input features, the algorithm moves the label position in which removes the label conflict or generates an alternative location (please refer to section 3.5).

To verify the adoptability of this phase, the number of label-feature conflicts before and after the sliding phase is summarized in Table 2. It is obvious that as the input features get larger, the set of the solution increases, too. For example, the total possible solution is $24^{3351}$ It is a huge number for 3351 features with 24 candidate locations. Thus, finding the optimal solution from such a number is a challenging task with 10000 iterations. As a consequence, many labels are placed in overlap with features for the first dataset. The iteration cycle is set to 10000 for all input maps, hence the possibility of finding the optimal solution from $24^{3351}$ is almost zero; in spite of this, our results are promising compared to MapLex and other algorithms. Sliding label position has further improved the obtained results, such as 51 label-feature conflicts being removed for the first dataset, and the score value decreased from 45 to 34. Moreover, as the table illustrates, when the input features get smaller, the number of label-feature conflicts reduces because the algorithm is able to explore the best solution with a lower number of iterations. The column named 'Better position' shows that the labels are replaced, but still, they are in conflict with features.

Table 2. Label replacement based on sliding model.

| Dataset | Eliminate label-feature conflicts | Better position | Previous score | New score |
|---|---|---|---|---|
| US counties | 51 | 32 | 45.28 | 34.01 |
| OSM | 10 | 3 | 16.35 | 12.31 |
| Ohio | 6 | 4 | 36.05 | 31.5 |
| Washington | 3 | 1 | 5.86 | 4.42 |

*4.3 Comparison Analysis with Other Algorithms*

Two critical components in an algorithm are computation efficiency and quality results. This study focused on reducing the computation time of map labeling and achieving a promising result. In order to evaluate the applicability of the proposed approach, the results are compared with (Deng et al., 2021; Lessani et al., 2021; Lu et al., 2019) algorithms. Three quality metrics are taken into consideration in comparison analysis: the execution time, the number of label-label and label-feature conflicts, and the quality of label positioning. The statistical results are presented in Tables (3-5).

Table 3. summarizes the results of MapLex, Parallel-MS, and the proposed approach on an OpenStreetMap dataset. Parallel-MS introduced a parallel algorithm for label placement (Lessani et al., 2021), yet its computation time can be a critical point besides having a higher number of label-feature conflicts. For example, the execution time for 814 features is over 118 minutes with 21 feature-label conflicts, though running in parallel. On the contrary, our approach is able to complete around 8 minutes with only 11 feature conflicts, and it is a huge improvement in terms of computation efficiency in addition to the reduction of label-feature conflicts. As the table illustrates, the lowest score value varies in both algorithms because of applying a different quality function. The OSM dataset is only labeled by (Lessani et al., 2021), and the proposed algorithm since (Deng et al., 2021; Lu et al., 2019) would take significantly a long time.

Tables (4) and (5) illustrate the results of (Deng et al., 2021; Lessani et al., 2021; Lu et al., 2019) algorithms and our method on two datasets (Ohio and Washington maps).

DDEGA-NM (Deng et al., 2021) presents a comprehensive study regarding multiple geographical feature labeling. Its computational time is unsatisfactory however. For instance, the Ohio dataset only includes 140 features, yet DDEGA-NM takes 3221 minutes to complete label positioning, and 40 labels are in conflict. On the other hand, Parallel-MS reduced the execution time, and the number of label-feature overlaps greatly. Nonetheless, the computation time is still unacceptable, which is roughly 46 minutes. On top of that, the obtained score verifies that most labels are positioned at their unpreferred location using DDEGA-NM and Parallel-MS algorithms. Similar results are achieved for the Washington map dataset. For a similar dataset, the running time is less than one minute in the serial version of the proposed algorithm, while the Parallel-MS algorithm took 20 minutes in parallel. To end that, our approach is able to achieve a promising solution not only in terms of label placement quality but also with less computation time.

Table 3. Statistical results of different algorithms on OSM dataset.

| Algorithms | Conflicts | | Lowest score | Iterations | Obtained score | Time (min) |
|---|---|---|---|---|---|---|
| | Feature | Label | | | | |
| MapLex | 37 | 0 | // | // | // | // |
| Parallel-MS | 21 | 0 | 2.04 | 10000 | 18.06 | 118.45 |
| Proposed | 11 | 0 | 4.52 | 10000 | 12.31 | 8.34 |

Table 4. Statistical results of different algorithms on Ohio State map.

| Algorithms | Conflicts | | Lowest score | Iterations | Obtained score | Time (min) |
|---|---|---|---|---|---|---|
| | Feature | Label | | | | |
| DDEGA-NM | 40 | 3 | 11.56 | 10000 | 63.04 | 3221.64 |
| Parallel-MS | 29 | 3 | 11.56 | 10000 | 39.26 | 45.98 |
| Proposed | 10 | 2 | 9.28 | 10000 | 27.87 | 3.51 |

Table 5. Statistical results of different algorithms on Washington State map.

| Algorithms | Conflicts | | Lowest score | Iterations | Obtained score | Time (min) |
|---|---|---|---|---|---|---|
| | Feature | Label | | | | |
| DDEGA | 37 | 0 | // | 300 | // | 3060 |
| DDEGA-NM | 22 | 0 | 1.56 | 10000 | 8.14 | 2166 |
| Parallel-MS | 12 | 0 | 1.56 | 10000 | 1.91 | 20.01 |
| Proposed | 6 | 0 | 4.41 | 10000 | 4.42 | 0.43 |

*4.4 Discussion*

All results based on four datasets verify the performance of the developed algorithm regarding the speed and label placement quality. For instance, the Washington dataset includes 71 multiple geographical features (please refer to Table 5), and we used three previous algorithms to label the features besides the proposed algorithm. According to comparison analysis, among 71 features, only 6 label-feature conflicts exist in our result, while the Parallel-MS algorithm obtained a result with 12 label-feature conflicts.

Moreover, the running time of the Parallel-MS algorithm is over 20 minutes though it utilizes multiple cores; yet, the computation time of the introduced algorithm is less than one minute though the algorithm uses one CPU for this dataset. As the table illustrates, the column named 'Lowest score' presents the optimization value of the established quality function for this algorithm; in other words, if all labels place at their most desirable location regardless of label-label conflict, the value of the quality function is equal to these values. On the other hand, the column termed 'Obtained score' shows the lowest score values achieved by the algorithm after the iteration cycle. As it can be observed, these values are greater than the values of the 'Lowest score' column due to the fact that if all labels place at their most preferred location, label-label conflict probably occurs. Since the weight value for label-label conflict is set greater than label-feature conflict in the quality function, thus, it is highly unlikely for the algorithm to get a score value equal to the optimization score value. This is true for all other datasets. As shown in the tables, the score value of our quality function is different than previous studies because of considering more quality metrics and different weight values for label-feature conflicts.

Figures 10 and 11 elucidate that in this algorithm, the performance of the optimization stage is enhanced as well. Population initialization and maintaining diversity in chromosomes are considered critical in evolutionary algorithms. On that account, one population is generated according to the best location of the label for each feature among 100 initial populations. This enables the algorithm to explore a satisfactory result in a lower number of the iteration cycle. As demonstrated in Figure 10(b), when the population is randomly initiated, the algorithm requires more iterations to achieve the same result as the nonrandom population; for example, the score values for the first iteration are 119 and 80 for random and nonrandom populations, respectively. Additionally, the lowest score is over 16 when the population is selected randomly. The score value is around 12 with the same number of iterations when the population is selected based on the best possible solution. Besides, all deployed CPUs share their best solution with each other during the iteration cycle, as presented in Figure 11. This increases the probability of identifying a promising solution within a few iterations.

Even though our algorithm significantly improved multiple geographical features label placement, yet some limitations exist and require further investigation: one label is assigned for each polyline regardless of considering the length of the line, and this leads to increase ambiguity; only those labels are allowed to move or to be replaced when they only intersect with one type of feature in the sliding stage; 24 candidate positions are generated for every line and point features, though it is possible to reduce the number of candidate locations for some features when the features are spares. These drawbacks may reduce the quality of label placement. Therefore, future researchers are encouraged to address these challenges and incorporate them into our algorithm.

## 5    Conclusion

The experimental results illustrate the practicality of the introduced algorithm that is able to achieve a promising solution with high-performance computing. This study intended to address the problem of multiple geographical feature label placement with regard to computation time reduction and improving label readability. The algorithm is developed based on a fixed position model and sliding model, attempting to further enhance the eligibility of map labeling. Furthermore, Message Passing Interface (MPI) is regarded to run the algorithm concurrently, aiming to reduce the overall computation time. The results reveal that considering these steps indeed enormously improves label positioning compare to other studies in addition to MapLex, as discussed in comparison analysis section. Thus, the proposed algorithm can be considered the basis of future studies.

To quantitatively evaluate label placement, a comprehensive quality function is introduced for the proposed algorithm. Additionally, one position is generated for those area features that are free of intersection with line and point features to save time in the candidate position phase and reduce the complexity of the problem in the optimization stage. To verify the applicability of the introduced algorithm, four real-world datasets were selected for the experimental analysis. The computation time for 71 mixed-types of feature is 43 seconds in the serial version of our algorithm, while Parallel-MS takes over 20 minutes despite using multiple cores, and the number of label-feature conflicts is 6 and 12, respectively. In addition, (Deng et al., 2021; Lessani et al., 2021; Lu et al., 2019) studies randomly generate the initial population in the optimization phase. However, our algorithm is able to initialize the population based on a better solution. Also, the implementation of communication among CPUs further enhances the performance of the optimization algorithm.

For future developments, with regard to computer graphic standardization, it would be appealing to investigate the effectiveness of the developed algorithm with more sophisticated multiple geographical features. Furthermore, the visualization and scale of the map have undergone significant changes in recent decades. Therefore, exclusive quality factors are essential to be included, such as multiple labels for a polyline and producing the candidate locations more accurately to decrease the possibility of label-feature conflict. Due to the fact that parallel processing is applicable in label placement, distributed computing frameworks, therefore, are experiential, and it is a new approach in the realm of parallel programming with high-performance computing. Thus, an interesting and pivotal future work is to customize the presented algorithm to label multiple geographical features.

**Data and codes availability statement**

The data that support the findings of this research is openly available in figshare repository [https://figshare.com/articles/dataset/Data/21597420].

**Disclosure statement**

No potential conflict of interest was report by the authors.

**Funding**

This research is supported by South Carolina SeaGrant (135400-22-60319).